\documentclass{elsart}
\usepackage{graphics}
\newcommand{\ie}{{\em i.e.} }
\newcommand{\sgn}{\mathop{\mathrm{sgn }}}

\begin{document}

\begin{frontmatter}
\title{Mobility and Reactivity of Discrete Breathers}

\author[saclay]{Serge Aubry} and \author[lyon]{Thierry Cretegny}
\address[saclay]{Laboratoire L\' eon Brillouin (CEA-CNRS), CE Saclay\\
91191-Gif-sur-Yvette Cedex, France\\ email: aubry@bali.saclay.cea.fr}
\address[lyon]{Laboratoire de Physique de l'Ecole Normale Sup\'erieure de Lyon, \\
CNRS URA 1325, 46 all\'ee d'Italie, 69007 Lyon, France\\  email: tcretegn@physique.ens-lyon.fr}

\begin{abstract}
Breathers may be mobile close to an instability threshold where the frequency of
a pinning mode vanishes. The translation mode is a marginal mode that 
is a solution of the linearized (Hill) equation of the breather which 
grows linearly in time. In some cases, there are exact mobile breather 
solutions  (found numerically), but these solutions have an infinitely 
extended tail which shows that the breather motion is nonradiative only
when it moves (in equilibrium) with a particular phonon field.
 
More generally, at any instability threshold, there is a marginal mode.
There are situations where excitations by marginal modes 
produce new type of behaviors such as the fission of a breather. We 
may also have fusion. This approach suggests that breathers (which 
can be viewed as cluster of phonons) may react by themself
or one with each others  as well as in chemistry for atoms and molecules, or
in nuclear physics for nuclei.
\end{abstract}
\end{frontmatter}

\section{Breathers by the Principle of Anticontinuity:
Brief Review of Current Developments}

The concept of nonlinear self-localization which is now emerging as ubiquitous in many
highly nonlinear models, has an old historical origin. Although exact localized and
time periodic solutions were already known for long in special integrable models
such as the sine-Gordon \cite{SCL73} or the Ablowitz-Ladik equation \cite{AL76}, these solutions
were nongeneric because they could not survive under most perturbations of the models.
Landau was probably the first physicist who pioneered the concept of nonlinear localization
when he found in 1933 that very generally a quantum electron strongly
coupled to a polarizable medium could localize in a potential well
created self-consistently and then form a polaron\cite{Lan33}. 
Actually, the concept of breathers appeared implicitly
in many fields in physics but without its generality and
terminology. As a unique example, it is straightforward to reinterpret
in terms of breathers the old well-known Josephson Junction (JJ)
effect between two superconductors.  In the undamped limit, it is
described by nothing but a Discrete Nonlinear Schr\"{o}dinger equation
with two sites only \cite{AFKO96}. Then, the presence of a breather is
just associated with a rotation of the phase difference between the
two superconductors that is the JJ effect (The same ideas extend to JJ
arrays \cite{FMMFA96,Flo97}).  Despite many precursive ideas, it was
only in 1988 that Takeno and Sievers clearly suggested that
intrinsically localized modes ({\it breathers}) in discrete anharmonic
systems should be quite general and robust solutions existing in many
nonlinear models \cite{ST88} (for a review see \cite{FW97}). We
believe that the discovery of this concept is a major achievement in
the recent years and should become the cornerstone of new important
development in Nonlinear Physics with applications to a wide number of
fields in Physics, Chemistry and Biology.

The first rigorous proof for the existence of breathers in a wide
class of models was given later \cite{MA94}. This proof was obtained
by considering first a limit where the model reduces to a discrete
array of uncoupled anharmonic oscillators. Then, breathers,
corresponding to one oscillator moving freely, trivially exist and can
be continued up to nonzero values of the coupling.  This uncoupled
limit which can be viewed as opposite to the integrable limit where
the system is harmonic and spatially continuous, was named
anti-integrable or anticontinuous \cite{AA90,Aub94}. This theory also
holds with aperiodic lattices which may be random, it holds for
coupled rotors (rotobreathers), for electrons coupled to anharmonic
oscillators (polarobreathers) etc\ldots \cite{Aub97}.

The basic principle of these proofs can be extended for proving the
existence of breathers in nonsymplectic models with dissipation
\cite{MKS95,SMK97}.  It can be also extended to anharmonic models
involving acoustic phonons for example to an extension of the
d-dimensional version of the model described in \cite{FCP97} which
couples linearly anharmonic optical variables with harmonic acoustic
variables \cite{AP97}. A different approach is proposed in
\cite{LSMK97} which introduces a new type of anticontinuous limit in
diatomic FPU chains where the mass ratio between the light and heavy
atoms goes to zero.  It can be extended to higher dimensions
\cite{MK97}.  These different extensions confirm how general is the
concept of discrete breather, and that they can persist as exact
solutions despite the increasing of the model complexity.

The same theory also predicts the existence of multibreather solutions
obtained by continuation from the solutions where several oscillators
move at the same frequency \cite{Aub97} and many of them are linearly
stable. They can be clusters of a few number of breathers and then are
still considered as breathers (e.g. 2-sites breathers). They can also
be extended over arbitrary infinite clusters. When the cluster covers
all the lattice sites, the multibreather becomes just an anharmonic
plane waves.  They have the property that they can transport energy by
phase torsion over arbitrarily complex pattern called {\it rivers}
\cite{CA97}. Such solutions can persist even when the lattice is {\it
random}.  In that case, it is interesting to point out that these
multibreather solution can transport energy while all the linear modes
are completely localized and cannot transport any.

These existence proofs can be also turned into an efficient method for
the practical calculation at the computer accuracy of any breather and
multibreather solutions \cite{MA96,CAT96,FMMFA96,CA97,JA97,FCP97} in
any model where they exist (including FPU chains).
   
At the anticontinuous limit, the breather quantization is trivial
\cite{Aub97}. Within the standard action-angle representation, the
operator corresponding to the action of the uncoupled anharmonic
oscillator at site $i$, can be written as $I_{i}=a^{+}_{i}a_{i}+1/2$
using standard creation and annihilation boson operators $a^{+}_{i}$
and $a_{i}$ (phonons).  The anharmonicity of the oscillator is
equivalent to the fact that $H(I_{i})$ is not a linear function of
$I_{i}$, that is the phonons are interacting. Then, a quantum breather
just appears as a bonded (or antibonded) cluster of phonons at a given
site and no phonon elsewhere.

When the oscillators are coupled with some coupling $C$ which has to
be small compared to the binding energies, a naive perturbation
calculation shows that these bonded states of phonons should persist
as narrow bands.  Their band width is proportional to $C^p$ where $p$
is the number of phonons involved in the quantum breather. When $p$ is
large, this band width becomes negligible or equivalently the breather
mass infinite\footnote{This is the expected behavior when the
classical breather is not mobile and when there is no accidental
resonances between different breather structures with the same
energy.}. Such behavior is indeed observed for the dimer problem
\cite{AFKO96} where the band width is replaced by a narrow splitting
between two levels and also for the trimer \cite{Fle97}.  If quantum
breathers are narrow bands of bonded bosons, quantum polaron could be
also viewed as narrow bands of a bonded state between an electron and
many bosons. Quantum Bipolarons are the same but with 2 electrons in a
singlet state bonded to a cluster of phonons.

\section{Breather Mobility by Marginal Modes}

It has been observed since several years that discrete breathers could
be mobile in some models \cite{HT92,SPS92} (for a review see
\cite{FW97})\footnote{Strictly speaking mobile breathers are no more
time periodic solutions and they require a specific definition which
we shall give later.}.  Mobile breathers for which the velocity can go
to zero (or at least become small), can be studied by continuity from
the immobile breathers.  This problem looks analogous to the study of
mobile kinks which can be often well described with a collective
coordinate as a massive particle moving in a periodic Peierls-Nabarro
(PN) potential.  Many numerically observed features for mobile
breathers at low velocity, are indeed reminiscent of those of moving
kinks and for that reason, it has been proposed that breather mobility
could be also related to the vanishing of some Peierls-Nabarro barrier
\cite{BP95}.

However, it turned out that this concept cannot be clearly defined
\cite{FW97} at least as a straightforward extension of the standard
concept.  The reason is that discrete breathers are not topologically
conserved objects, and belong to a family of solutions the internal
energy (and frequency) of which can vary continuously. If a PN energy
barrier could be defined with a separatrix in the phase space, it
could be believed that the breather could nevertheless overcome this
barrier by tuning its frequency and releasing some of its internal
energy. This point has been discussed in details \cite{FW95}. This
point of view is not totally convincing because it is not discussed
whether this internal energy could be released.

We proposed a consistent definition for a PN barrier but for the
action of the breather instead of its energy \cite{Aub96}. This
approach could explain the formation of {\em intermediate} breathers
which correspond to the intermediate extrema (\ie not centered on a
lattice site nor in the middle of a bond) of the Peierls-Nabarro
action.  They are indeed numerically observed in many cases and shall
be discussed in another publication \cite{CMA97}.  We shall see here
that an {\it effective} PN barrier in energy could be nevertheless
defined empirically.

Beside the vanishing of its PN energy barrier, the mobility of a kink
is also related to the vanishing of the frequency of a pinning (or
translation) mode.  Breathers also may have internal modes which are
spatially localized.  Unlike the PN barrier, they have an unambiguous
definition which hold as well for kinks. These modes can be used for
testing the breather mobility.  When a breather is (almost) freely
mobile, there are small perturbations (namely kicks on the {\it
translation mode}) which induce its motion at slow velocity.  Unlike
most linear perturbations which grows exponentially in time (for
unstable modes) or remain bounded (for stable modes), such a
perturbation grows linearly in time.  We call such kind of
perturbations {\it marginal modes}.  In the limit of a zero velocity,
the ideal breather trajectory tends to become a continuum of breather
solutions (if it exists). As a result, there are also perturbations
which remain bounded in time and thus do not put the breather into
motion.  They correspond to a breather pinning mode at zero frequency.

For investigating pinning and marginal modes, let us consider as an
example, the standard discrete Klein-Gordon chain with Hamiltonian:

\subsection{The Klein-Gordon Chain: Notations}
\begin{equation}
	{\bf H} = \sum_n \left(\frac{p_n^2}{2} + V(u_n) + \frac{C}{2}
	(u_{n+1}-u_n)^2 \right) \label{hamiltKG}
\end{equation}
which consists of anharmonic oscillators with potential $V(x)$ and
mass unity coupled by a nearest neighbor harmonic coupling
with constant $C$.  Let us consider a time reversible breather
solution with period $t_b$ of the dynamical equation
\begin{equation}
\ddot{u}_n + V^{\prime}(u_n) - C (u_{n+1}+u_{n-1}-2 u_n) =0.
\label{dynam1} 
\end{equation} 
The linearized equation which
determines the ``harmonic'' modes of the breather
\begin{equation}
\ddot{\epsilon}_n + V^{\prime \prime}(u_n(t)) \epsilon_n - C
(\epsilon_{n+1}+\epsilon_{n-1}-2 \epsilon_n) =0
\label{linear1} 
\end{equation} 
is similar to a linear discrete
Schr\"{o}dinger equation but with a time periodic potential
$V^{\prime \prime}(u_n(t))$ with period $t_b$.  This equation
always has the trivial solution $\epsilon_n(t)= \dot{u}_n(t)$
(phase mode).

Integration over a period of time $t_b$ of this equation determines
the Floquet matrix ${\bf F}$ which relates linearly
$\{\epsilon_n(t_b),\dot{\epsilon}_n(t_b)\}= {\bf F}\{\epsilon_n(0),
\dot{\epsilon}_n(0)\} $ to its initial conditions
$\{\epsilon_n(0),\dot {\epsilon}_n(0)\}$. This matrix is $2N \times
2N$ for a system with $N$ sites.  It is symplectic which implies that
if $\lambda$ is an eigenvalue then $\lambda^*$, $1/\lambda$ and
$1/\lambda^*$ are also eigenvalues. The linear stability of the
breather requires that there are $N$ pairs of eigenvalues $\e^{\pm
i\theta_{\nu}}$ are on the unit circle (we choose for convenience $0
\leq \theta \leq \pi$).  The phase mode corresponds to a degenerate
pair at $\theta=0$.  This Floquet matrix is also connected to the phonon
scattering by the breather \cite{CAF97}.

The linearly stable mode associated with an eigenvalue $ \e^{i\theta_\nu}$
is time quasiperiodic and exhibits the series of frequencies
$\omega_{\nu}= (\frac{\theta_{\nu}}{2\pi}+n) \omega_b $ with $n$
integer. Thus, its frequency is defined modulo $\omega_b$. We choose
the determination $n=0$ and $0\leq \omega_{\nu} \leq \omega_{b}/2$.

\subsection{Linear Stability and Spectrum of the Second Variation of the 
action} It has been shown in \cite{Aub97} that it is also convenient
 to study the equation
\begin{equation}
	\ddot{\epsilon}_n + V^{\prime \prime}(u_n) \epsilon_n - C
	(\epsilon_{n+1}+\epsilon_{n-1}-2 \epsilon_n) = E \epsilon_n
	\label{linear2} 
\end{equation}
which allows one to explicit the Krein theory of bifurcations in
simple terms and in addition, yields more informations.  This is
nothing but the eigenequation of the matrix of the second variation of
the action expanded around the breather solution which by definition
is one of its extrema. This is also the (full) Newton matrix involved
in the breather continuation \cite{Aub97}.

\begin{figure}[tbp]
\centering
\resizebox{\textwidth}{!}{\includegraphics{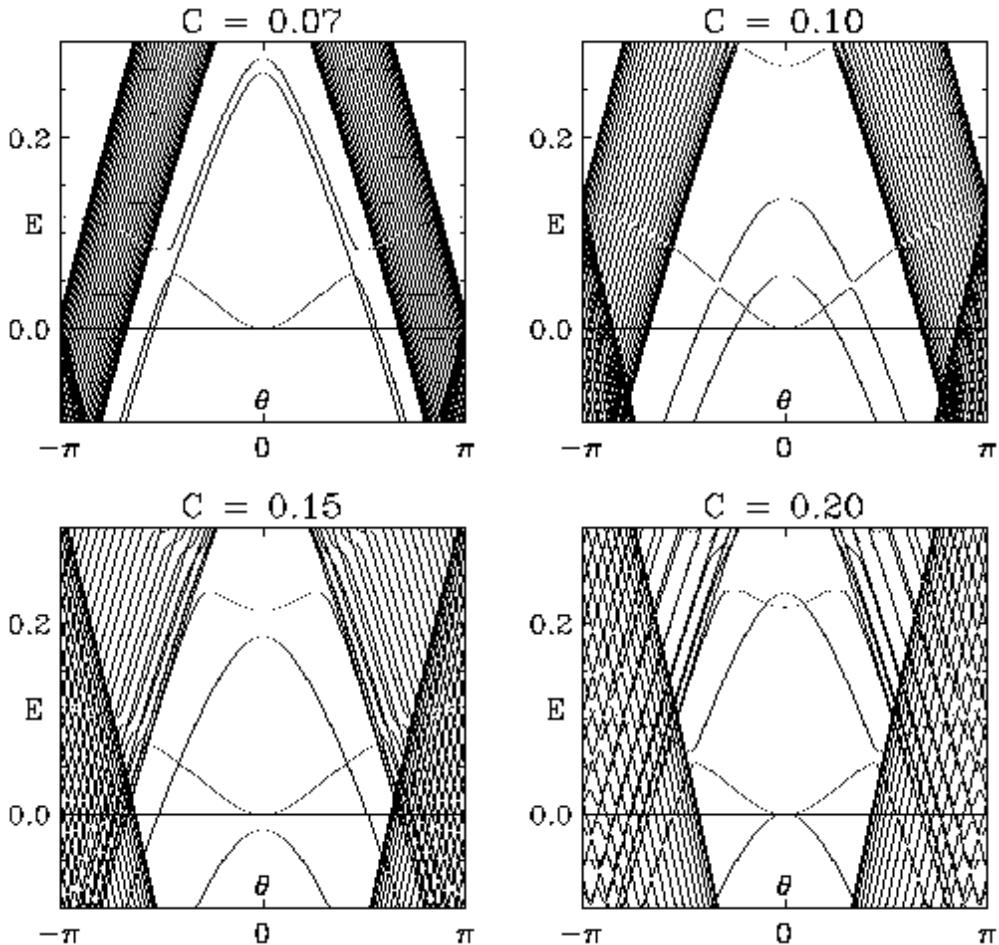}}
\caption{ Bands $E_{\nu}(\theta)$ defined by eqs.(\ref{linear2}) and
(\ref{Bloch}) calculated for the single breather at frequency
$\omega_b=0.75$ for the KG chain (\ref{hamiltKG}) with the Morse
potential $V(x)=\half (1-\e^{-x})^2$ and $C=0.07, 0.1, 0.15, 0.2$.  }
\label{fig1}
\end{figure}

Let us make a brief review of its properties.  The time periodicity of
$V^{\prime \prime}(u_i(t))$ implies that the eigenvalues
$E_{\nu}(\theta)$ of eq.(\ref{linear2}) forms bands indiced by
$\nu$. They are $2\pi$ periodic and symmetric functions of the
parameter $\theta$ defined modulo $2\pi$ and the corresponding
eigenstates fulfills the Bloch condition
 
\begin{equation}
	\epsilon_{\nu,n}(t+t_b, \theta) = \e^{i \theta t/t_b}
	\epsilon_{\nu,n}(t,\theta), \ \ \ \ \ n = 1, \ldots, N
 \label{Bloch} 
\end{equation}
(see fig.\ref{fig1} for an example of numerical calculation of the
bands $E_{\nu}(\theta)$ of eq.(\ref{linear2})).

The eigenvalues $\e^{\pm i \theta_{\nu}}$ of the Floquet matrix ${\bf
F}$ which are on the unit circle, are determined by the set of
intersections $\theta_{\nu}$ of the bands $E_{\nu}(\theta)$ with the
axis $E = 0$. The Krein signature of a pair of conjugated eigenvalues
$\e^{\pm i\theta_{\nu}}$ is defined as\cite{AA68}
\begin{equation}
\kappa(\theta_\nu) = \sgn \left( i \sum_n \epsilon_{\nu,n}
\dot{\epsilon}_{\nu,n}^* - \epsilon_{\nu,n}^* \dot{\epsilon}_{\nu,n}
\right) 
\end{equation} 
which is a conserved quantity due to the
symplecticity of the dynamics. According to the Krein theory, two
pairs of eigenvalues $\e^{\pm i\theta_\nu}$ and $\e^{\pm i\theta_\mu}$
which collide on the unit circle may lead to an instability only if
their signature is different. This criterion was reinterpreted in
\cite{Aub97}, where it has been shown that the Krein signature of the
eigenvalues $\e^{\pm i\theta_{\nu}}$ is the opposite sign of the slope
$\d E_{\nu}(\theta)/\d \theta$ at $\theta=\theta_{\nu}$.  Thus, it is
clear that a bifurcation can occur only if the two eigenvalues belong
to the same band (c.f. fig.\ref{fig2}). As a consequence, the slopes of
the band (or the Krein signatures) at $\theta_\nu$ and $\theta_\mu$
have to be different.

For a stable breather in a lattice with $N$ sites, there are $N-1$
bands intersecting the axis $E=0$ and one corresponding to the phase
mode which is tangent to this axis at $\theta=0$. When the system is
infinite, there is a continuum of bands (see fig.\ref{fig1}) which can
be easily calculated because it corresponds to the spectrum of the
system without breather ($u_{i}(t) \equiv 0$ and $V^{\prime \prime}
(u_{i}(t))= \omega_{0}^{2}$) .  There is at least one isolated band
corresponding to the phase mode and possibly some other isolated bands
which correspond to spatially exponentially localized modes.

\subsection{Marginal Modes: Existence Proof}  \label{subsection2}

\begin{figure}[tbp]
\centering 
\includegraphics{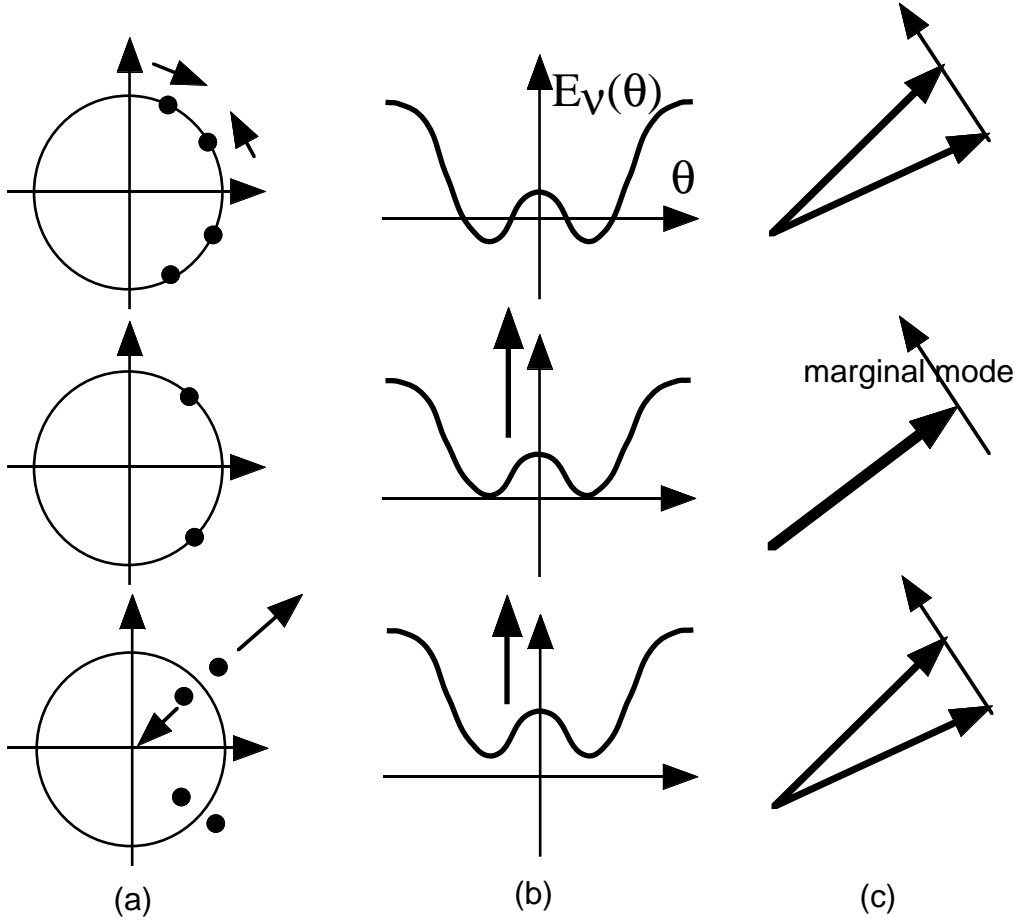} 
\caption{Schemes showing as an example the
evolution of four eigenvalues on the unit circle (a), the
corresponding band shape of the matrix of the second variation of the
action(b), the corresponding eigenvectors of the Floquet matrix for a
Krein crunch before, at and after the bifurcation (c).} \label{fig2}
\end{figure}

When the breather becomes linearly unstable, one of these bands
$E_{\nu}(\theta)$ moves and looses its intersection (see
fig.\ref{fig1} between $C = 0.1$ and $C = 0.15$, fig.\ref{fig2} and
\cite{Aub97} for details).  This appears on the unit circle as a
collision between two eigenvalues $\theta_{\nu}$ and $\theta_{\mu}$.
At the bifurcation, we have three possible situations whether the
curve is tangent to $E=0$ either at $\theta=0$, or at $\theta=\pi$ or
at $\theta=\theta_0 \neq 0$ or $\pi$ (Krein crunch). The scheme on
fig.\ref{fig2} shows an example for a Krein crunch.

Another important result which comes out readily from our band
analysis is that the two eigenmodes associated with the eigenvalues
$\e^{i \theta_{\nu}}$ and $\e^{i \theta_{\nu}^{\prime}}$ are colinear
at the bifurcation (see fig.\ref{fig2}).  Thus, the Floquet matrix
${\bf F}$ looses either one ($\theta=0$ or $\theta=\pi$) or two
eigenvectors ($\theta=\theta_0 \neq 0$ or $\pi$).  As a consequence,
the space generated by the eigenvectors of ${\bf F}$ is not the whole
space. We show that the missing modes are just {\it marginal modes}
which grows linearly in time.

Let us prove that at each bifurcation where there is a curve
$E_{\nu}(\theta)$ tangent from above (or from below) to the line $E=0$
at $\theta=\theta_0$, eq.(\ref{linear1}) exhibits a marginal mode.

The eigensolutions of eq.(\ref{linear2}) associated with the
eigenvalue $E_{\nu}(\theta)$ which fulfills the Bloch condition
(\ref{Bloch}) can be written with the form
$\epsilon_n^{\nu}(t,\theta)= \e^{i \theta t/t_b}
\chi_n^{\nu}(t,\theta)$ where $\chi_n^{\nu}(t,\theta)$ is time
periodic with period $t_b$. We assume for example that
$E_{\nu}(\theta)$ is tangent to $E=0$ from above. Then for $E>0$ small
enough, there are two values $\theta_1(E)$ and $\theta_2(E)$, such
that \begin{itemize} \item $E_{\nu}(\theta_1)=E$ and
$E_{\nu}(\theta_2)=E$ \item $\lim_{E \rightarrow
0}\theta_1(E)=\theta_0$ and $\lim_{E \rightarrow
0}\theta_2(E)=\theta_0$.  \end{itemize} Since the combination
$$\frac{\epsilon_n^{\nu}(t,\theta_1(E))-\epsilon_n^{\nu}(t,\theta_2(E))}
{\theta_1(E)-\theta_2(E)}$$ is also an eigensolution solution of
eq.(\ref{linear2}), it comes out that for $\theta = \theta_0$,
\begin{equation} \frac{\partial \epsilon_n^{\nu}(t,\theta)}{\partial
\theta} = \e^{i \theta t/t_b} \left(i \frac{t}{t_b}
\chi_n^{\nu}(t,\theta) + \frac{\partial
\chi_n^{\nu}(t,\theta)}{\partial \theta}\right) \label{lindiv}
\end{equation} is a solution of eq.(\ref{linear1}) which diverges in
time proportionally to $t$ and thus is a marginal mode ($\partial
\chi_n^{\nu}(t,\theta)/ \partial \theta$ is time periodic with period
$t_b$).

Thus, at the bifurcation, the Floquet matrix ${\bf F}$ exhibits a
stable eigenmode $\left(\{\epsilon_n^{\nu}(0,\theta_0)\},\{
\dot{\epsilon}_n^{\nu} (0,\theta_0)\} \right)$ for the eigenvalue
$\e^{i\theta_0}$ associated with a marginal mode
$$\left(\{\frac{\partial \epsilon_n^{\nu}(0,\theta_{0})}{\partial
\theta}\}, \{ \frac{\partial
\dot{\epsilon}_n^{\nu}(0,\theta_{0})}{\partial \theta}\} \right)$$
which grows linearly in time. When $\theta_0 \neq 0$ and $\pi$, the
complex conjugate eigenvectors have the same properties.  This
property is associated with the fact that the $2 \times 2$ Floquet
matrix restricted to the subspace determined by the two eigenvectors
which becomes colinear at $\theta=\theta_0$, has degenerate
eigenvalues but only one eigenvector and thus is not
diagonalizable. The marginal mode is not uniquely defined. It could be
combined arbitrarily with the associated eigenmode but this is just
equivalent to change the origin of time.

As we pointed above, there is always a band $E_{\nu}(\theta)$ tangent
at $E=0$ for $\theta_0=0$. Equation (\ref{linear1}) exhibits a phase
mode ($\epsilon_n(t)= \dot{u}_n(t)$) for any breather solution and an
associated marginal mode which diverges linearly in time. The latter
can be also obtained directly by derivation of $u_n(t)= g_n(\omega_b
t, \omega_b)$ with respect to $\omega_b$ and thus does not represent
physically a real instability.  Since an excitation of this mode
produces a small change in the breather frequency, it is called growth
mode.

\subsection{Kicking a Breather}

Turning back to the search of pinning mode and its associated marginal
mode, we have to look for the breather bifurcations at $\theta_0=0$. A
systematic analysis of the Floquet matrix of the breathers can be
accurately and easily done numerically with the new methods developed
in \cite{MA96}. It yields many bifurcations and some of them are at
$\theta=0$. We just have to test numerically the effect of a breather
perturbation in the direction of their marginal mode\footnote{Of
course, perturbations in the direction of the growth mode, should not
be excited for producing mobile breathers since its effect is just to
change its frequency.}. 

In the case of a spatially symmetric breather, the eigenmodes of
eq.(\ref{linear2}) are either spatially symmetric or antisymmetric. As
a result the symmetric and antisymmetric branches can intersect one
with each other without interaction ({\em c.f. }  fig.\ref{fig1}). The
pinning mode which corresponds to a small translation of the breather,
has to break its spatial symmetry and should be searched as a
spatially antisymmetric mode.

An example has been studied in detail for the KG model
(\ref{hamiltKG}) with a double well potential $V(x)= \quart (x^2-1)^2$
\cite{CAT96}. More extensive numerical studies will be reported in
\cite{CMA97}. We shortly recall the main findings and refer the reader
to this reference for the illustrating figures.

For the double well potential and a single breather at frequency
$\omega_b= 2\pi/6 < \omega_0 = \sqrt{2} $, the numerical analysis
reveals that at $C=C_c \approx 0.5888$, there is a bifurcation at
$\theta=0$ concerning a spatially localized and antisymmetric
eigenmode. There is a pair of isolated eigenvalues $\e^{\pm i
\theta_1(C)}$ of the Floquet matrix ${\bf F}$ which go to unity for $C
\rightarrow C_c$. The time reversibility of the breather solution
implies that the pair of eigenvectors are complex conjugate and
one is the image of the other by time reversibility. Thus they have
the form ${\bf V}_\pm 
=\left(\{\delta_n(C)\}, \pm i\{\gamma_n(C)\}\right)$ where the
position component is real and the velocity component is purely
imaginary. When $C \rightarrow C_c$, $\gamma_n(C_c)=0$. The
pinning mode $\left(\{\delta_n(C_c)\}, \{0\}\right)$ only concerns
breather perturbations on the position of the particles.  The marginal
mode which is both time antisymmetric and spatially antisymmetric, is
obtained as the limit of the normalized vector $\lim_{C\rightarrow
C_c} {\bf \hat{V}}(C)$ where ${\bf \hat{V}}(C)= \frac{{\bf
V}(C)}{\|{\bf V}(C)\|_{2}}$ with ${\bf V}(C)=\left(\{0\},
\{\gamma_n(C)\}\right)$. It contains only velocity components.

When a small initial perturbation with no component on the marginal
mode is added to the breather, it does not move uniformly but only
oscillates.  On the opposite, it will move when the initial
perturbation has a component on the marginal mode $\lambda {\bf
\hat{V}}(C_{c})$. The breather motion appears to be the most perfect
(that is the most free of phonon radiation) when this perturbation is
a pure marginal mode.  A small perturbation $\lambda {\bf
\hat{V}}(C_{c})$ added to the initial conditions of the breather at
$C=C_{c}$, grows linearly in time as expected. Actually, the breather
moves slowly and the motion is uniform and persists over very long
time and many lattice spacing with almost no energy dissipation. Let
us emphasize that this breather mobility is not related to a large
spatial extension of the breather as it is usually believed at least
for mobile kinks.

This initial perturbation only concerns the initial velocity of the
particles, which at $t=0$ are zero for the unperturbed breather. Since
$\hat{\bf V}(C_{c})$ is normalized, the kinetic energy added
to the breather by this initial kick is just $\half \lambda^{2}$.  It
is found that the resulting velocity of this breather ${\it v}$ is
proportional to the amplitude $\lambda$. As a result, an effective
mass $m^{*}$ can be defined by the equality $\half \lambda^{2}=\half
m^{*}{\it v}^{2}$.

\begin{figure}
\includegraphics{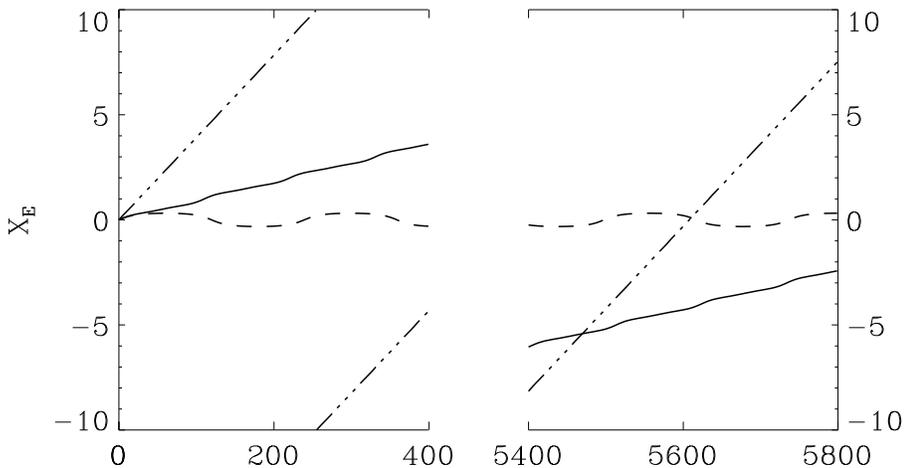}
\caption{Evolution of the center of energy (modulo 20) of a breather
perturbed 
along its translation mode with various amplitudes: $\lambda <
\lambda_c$ (dashed line), $\lambda \approx \lambda_c$ (full line) and
$\lambda > \lambda_c$ (dash-triple dotted line). The on-site potential
is the Morse potential, $C = 0.129$ and the frequency of the breather is
$\omega_b = 0.80$. The unit of time is the period of the perturbed
breather.}
\label{fig:evolX}
\end{figure}

When $C$ is close to but smaller than $C_{c}$, the breather does not
move but oscillates if the amplitude $\lambda$ of the initial
perturbation $\lambda \hat{\bf V}(C)$ is too small. This is not
surprising because this perturbation does not correspond to a marginal
mode but to an ordinary mode although its frequency is low. However,
there exists a quite well-defined value $\lambda_{c}(C)$ such that
when $\lambda > \lambda_{c}(C)$, the breather starts to move.  Just
beyond the threshold the motion is non uniform, similarly to a rotating
pendulum near the separtrix. For larger kicks, the breather moves almost
non-dissipatively over very long distances.
An example is presented on fig.\ref{fig:evolX} where a narrow
linearly stable Morse breather is made mobile through a kick along its
pinning mode.
In some cases, the breather may dissipate phonon radiation and stop
after some time.  As a result, the
determination of this mobility threshold cannot be determined with a
high accuracy but its order of magnitude is nevertheless physically
significant: $1/2 \lambda_{c}^{2}$ can be interpreted as the {\it
Peierls Nabarro energy barrier} which must be provided to move the
breather. It is nonzero for $C<C_{c}$ and vanishes at $C=C_{c}$. It is
undefined above (but could be defined from another intermediate
breather \cite{CMA97}).

When $C$ becomes larger than $C_{c}$, the breather is unstable and in
principle a small perturbation $\lambda \hat{\bf V}(C)$ initially
diverges exponentially. However, while the amplitude of this
instability is weak enough, the mobility usually remains very good.

There are models for which the single breathers do not exhibit any
instability threshold. Thus, no pinning mode can be found for example
in the KG chain (\ref{hamiltKG}) with the quartic potential $V(x)
=\half x^{2}+ \quart x^{4}$. All the numerical tests\cite{CAT96} we
did by kicking these breathers by reasonably small initial
perturbations, were unsuccessful to move the breathers even by few
lattice spacings.  However, a good mobility can be obtained when the
breather frequency approaches the phonon band edge where the size of
the breather diverges.  This is not surprising because it is
well-known that continuous models which do have exact propagating
localized solutions (e.g. the Nonlinear Schr\"{o}dinger equation),
provides good approximation of the model in that limit.

In summary, highly mobile breather (which are not spatially very
extended), may be found at the instability thresholds of the immobile
breather.  The numerical investigation of the spectrum and the
eigenvalues of the Floquet matrix ${\bf F}$ allows one to discover
systematically mobile breathers which are not spatially extended. The
method has been applied successfully in various models including FPU
chain and should work in principle for models in higher dimension.  It
has also been applied for finding {\em mobile} and {\em small}
bipolarons in the Holstein-Hubbard model \cite{PA97}.

\section{Exact Mobile Breathers: Loop Dynamics}

Although the above method allows one to find breathers which moves
with almost no energy dissipation, it is not clear whether there
exists exact solutions corresponding to a moving breather.

Let us assume that we have an exact mobile breather solutions with
frequency $\omega_{b}$ moving at the velocity ${\it v}$. After a time
$T=1/{\it v}$, the breather has moved by one lattice spacing, but
since this time is generally incommensurate with the period
$t_{b}=2\pi/\omega_{b}$ of the breather, the breather phase has been
rotated by an angle $\alpha= T \omega_{b}$ modulo $2\pi$,
incommensurate with $2 \pi$. After a time $nT$ such that $n T
\omega_{b} \approx 0$ modulo $2\pi$, the breather returns to an almost
identical configuration shifted by $n$ lattice spacing.  Therefore a
moving breather should be viewed as a moving loop in the phase
space. Its trajectory is the set of trajectories generated by the
initial conditions with all possible phases.  Such a dynamics can be
described formally.

Considering a dynamical system with Hamiltonian ${\bf
H}(\{p_{i},u_{i}\})$ where $u_{i}$ and $p_{i}$ are conjugate variables
and Hamilton equations \begin{equation} \dot{p}_i = -\frac{\partial
{\bf H}}{\partial u_i} \hspace{3cm} \dot{u}_i = \frac{ \partial {\bf
H}}{\partial p_i}.  \label{dyn} \end{equation}

We associate with this system, a dynamical system of loops
$(\{p_{i,x},u_{i,x}\})$ where the coordinates depends not only on $i$
and but also on extra continuous variable $x$ on a 1-torus. For a
given period $t_b$, we have 
\begin{equation}
(\{p_{i,x+t_b}(y),u_{i,x+t_b}(y)\}) = (\{p_{i,x}(y),u_{i,x}(y)\})
\end{equation} 
for any $y$ which represents a (fictitious) time.  Its
trajectories extremalize the extended action 
\begin{equation}
\mathcal{A} = \int dy \left(\int_0^{t_b} dx \sum_i
p_{i,x}(y)(\frac{\partial u_{i,x}}{\partial x}+ \frac{\partial
u_{i,x}}{\partial y})- {\bf H}(\{p_{i,x}(y),u_{i,x}(y)\}) \right)
\label{actloop} 
\end{equation}
which yields the extended Hamilton equations 
\begin{equation} \frac{\partial p_{i,x}}{\partial
x}+\frac{\partial p_{i,x}} {\partial y} = - \frac{\partial {\bf
H}}{\partial u_{i,x}}, 
\qquad 
\frac{\partial u_{i,x}}{\partial
x}+\frac{\partial u_{i,x}}{\partial y} =\frac{\partial {\bf
H}}{\partial p_{i,x}}.  \label{dyna} 
\end{equation} 
The corresponding Hamiltonian for this loop dynamics is 
\begin{equation}
\mathcal{H_L}(y) = \int_0^{t_b} dx \left(\sum_i p_{i,x} \frac{\partial
u_{i,x}}{\partial x} - {\bf H}(\{p_{i,x},u_{i,x}\})\right)
\label{hamilloop} \end{equation} 
that is the action of the loop where
$x=t$ is taken as the time in the initial model. For a solution of
(\ref{dyna}), this effective Hamiltonian is independent of the
fictitious time $y$.  Time periodic solutions with period $t_b$ for
the single time dynamics (and in particular the breathers) correspond
to extrema of (\ref{hamilloop}) and are fixed points for this loop
dynamics.

\begin{figure}[tbp] \centering
\includegraphics{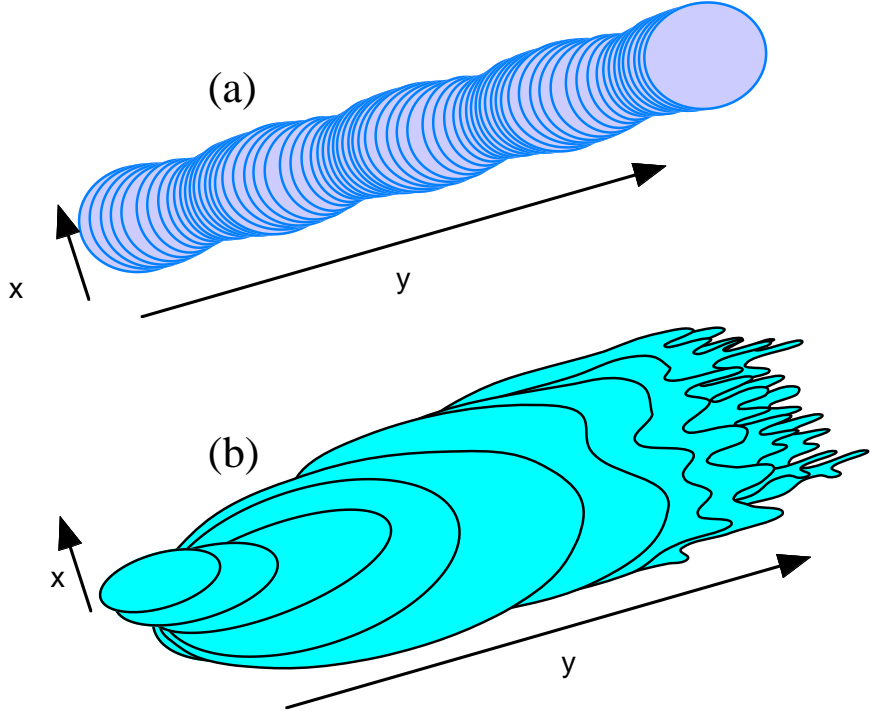} \caption{Scheme of a Loop Dynamics : (a)
represents the evolution in phase space of a loop which corresponds to
an exact mobile breather.  (b) the same for an arbitrary loop which
{\em mixes} in the phase space} \label{fig3} \end{figure}

An initial loop evolves in the phase space and in general if the
system is mixing it will spread densely over the phase space (see
fig.\ref{fig3}).  A mobile breather corresponds to a special solution
of eqs.(\ref{dyna}) which returns to an equivalent configuration apart
a space translation. It fulfills the skew periodicity condition
\begin{equation} (\{p_{i+1,x}(y+T),u_{i+1,x}(y+T)\}) =
(\{p_{i,x}(y),u_{i,x}(y)\}) \label{skewperiod} \end{equation} Let us
note that if $\{p_{i,x}(y),u_{i,x}(y)\}$ is a solution of
eqs.(\ref{dyna}), then for $a$ and $b$ arbitrary,
$\{p_{i,ax+(1-b)y}(by+(1-a)x),u_{i,ax+(1-b)y}(by+(1-a)x)\}$ is also
solution of the same equations. This change of variable does not
change at all the corresponding set of real trajectories. However, our
definition for the loop dynamics requires that the new solution be
periodic with respect to $x$, which implies $b=1$.  The new period
with respect to $x$ which is $t_{b}/a$, can be arbitrary.  In some
sense, this is not surprising because strictly speaking the frequency
of a mobile breather cannot be defined.  However, it is reasonable
that the loop motion in the phase space be as slow as possible. $a$
can be optimized for that and a possible criteria is that
$$\int_{0}^{T} dy \sum_{i}\left( (\frac{\partial p_{i}}{\partial
y})^2+ (\frac{\partial u_{i}}{\partial y})^2 \right)$$ be minimum.

If there is a solution of eqs.(\ref{dyna}) which fulfills
eq.(\ref{skewperiod}) for a velocity ${\it v}=1/T$, the energy of the
initial loop ${\bf H}(\{p_{i,x}(0),u_{i,x}(0)\})$ has to be constant
(i.e. independent of $x$).  It is convenient to write this solution
with hull functions $h_{i}(\xi,\eta,{\it v})$ and $g_{i}(\xi,\eta,{\it
v})$ which are $2\pi-$periodic with respect to $\xi$, and 1-skew
periodic with respect to $\eta$ (that is $h_{i+1}(\xi,\eta +1,{\it
v})=h_{i}(\xi,\eta,{\it v})$ and $g_{i+1}(\xi,\eta+1,{\it
v})=g_{i}(\xi,\eta,{\it v})$).  Then, we have
$(\{p_{i,x}(y),u_{i,x}(y)\})=(\{h_{i}(\omega_b x,{\it v} y,{\it v}),
g_{i}(\omega_b x,{\it v} y,{\it v})\})$.  Eqs.(\ref{dyn}) have a
solution with the form 
\begin{equation} p_i(t) = h_i(\omega_b t
+\alpha, {\it v} (t + \beta), {\it v}) \qquad u_i(t) = g_i(\omega_b t
+\alpha, {\it v} (t + \beta), {\it v}) \label{hull} 
\end{equation}
where $\alpha$ and $\beta$ are arbitrary phases.

Taking condition (\ref{skewperiod}) into account, we have
$h_{n}(\xi,\eta,{\it v}) = h_{0}(\xi,\eta-n,{\it v})$ and
$g_{n}(\xi,\eta,{\it v})= g_{0}(\xi,\eta-n,{\it v})$ which shows that
a unique pair of hull functions $h_{0},g_{0}$ is sufficient for
describing the whole breather motion. This new form turns out to be
identical to those proposed by S. Flach for a mobile breather
\cite{Fla97}.

We have no mathematical proof that such exact solution could exist,
but this problem could be approached numerically with the same methods
as those used for calculating non time reversible multibreathers
\cite{CA97}. It consists in the application of a modified Newton
method for finding fixed points of the Poincar\'e map ${\bf
T}_{L}:\{p_{i,x}(0), u_{i,x}(0)\} \rightarrow
\{p_{i+1,x}(T),u_{i+1,x}(T)\}$ defined by eqs.(\ref{dyna}) where the
period $t_{b}$ and pseudoperiod $T$ are given parameters.  It is clear
that as in \cite{CA97}, the Newton matrix is noninvertible at the
fixed point (if any). We know however that these degeneracies are associated
with the two arbitrary phases of eq.(\ref{hull}).  As for nontime
reversible breathers, this problem can be overcome by using a
technique of singular value decomposition which in some sense, is
equivalent to discard in the Newton matrix, the subspace associated
with the two eigenvalues which are small.  This method could converge
to an exact solution, if one chooses as initial solution a good
approximation of a mobile breather obtained by a marginal mode
excitation. This general method has not yet been implemented.

\begin{figure}[tbp] \centering
\includegraphics{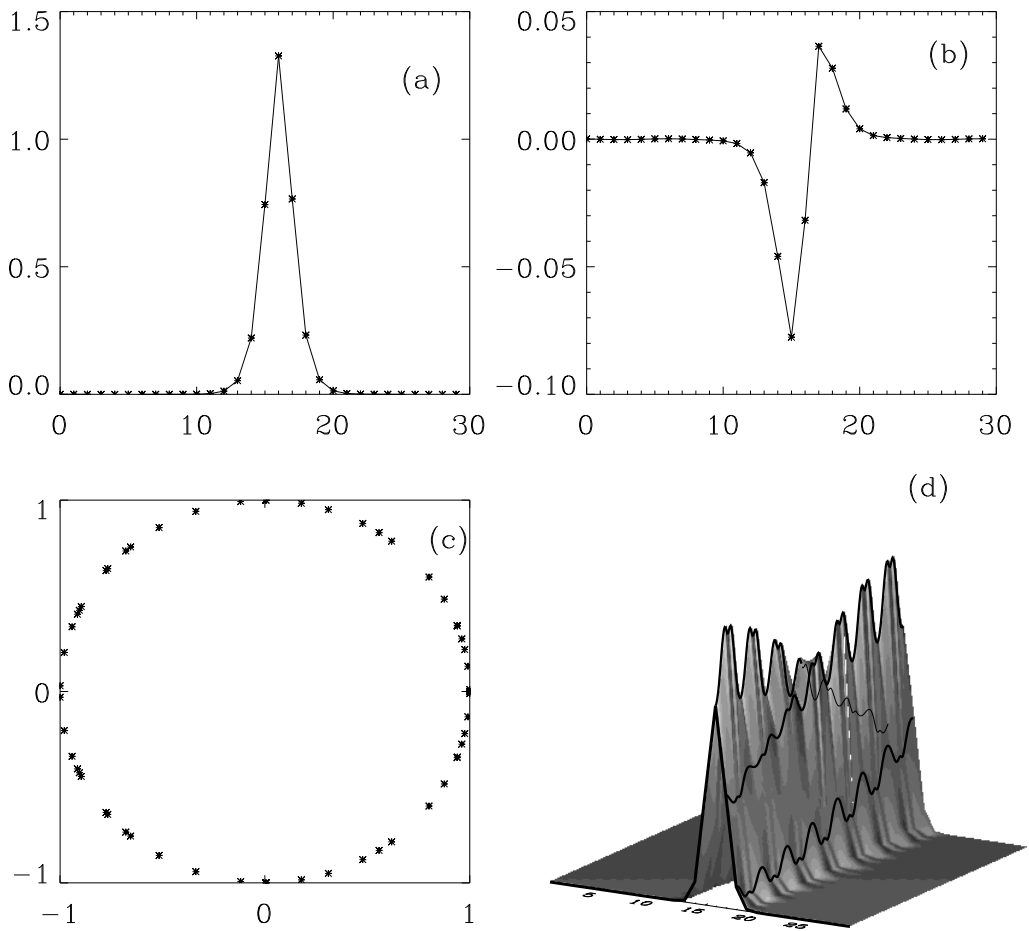} 
\caption{Numerically Exact Mobile
Breather obtained in the KG chain with the Morse potential at
$C=0.147$. The time needed to move over one lattice spacing is $T = 52.6$.
Profiles of the initial positions (a) and velocities (b). (c)
Distribution on 
the unit circle of the extended Floquet matrix ${\bf F}_{e}$.
This solution is linearly stable. (d) Space-time representation of the
energy density of the breather over one periode (there are $7$
oscillations). } \label{fig4a} 
\end{figure}
\begin{figure}[tbp] \centering
\includegraphics{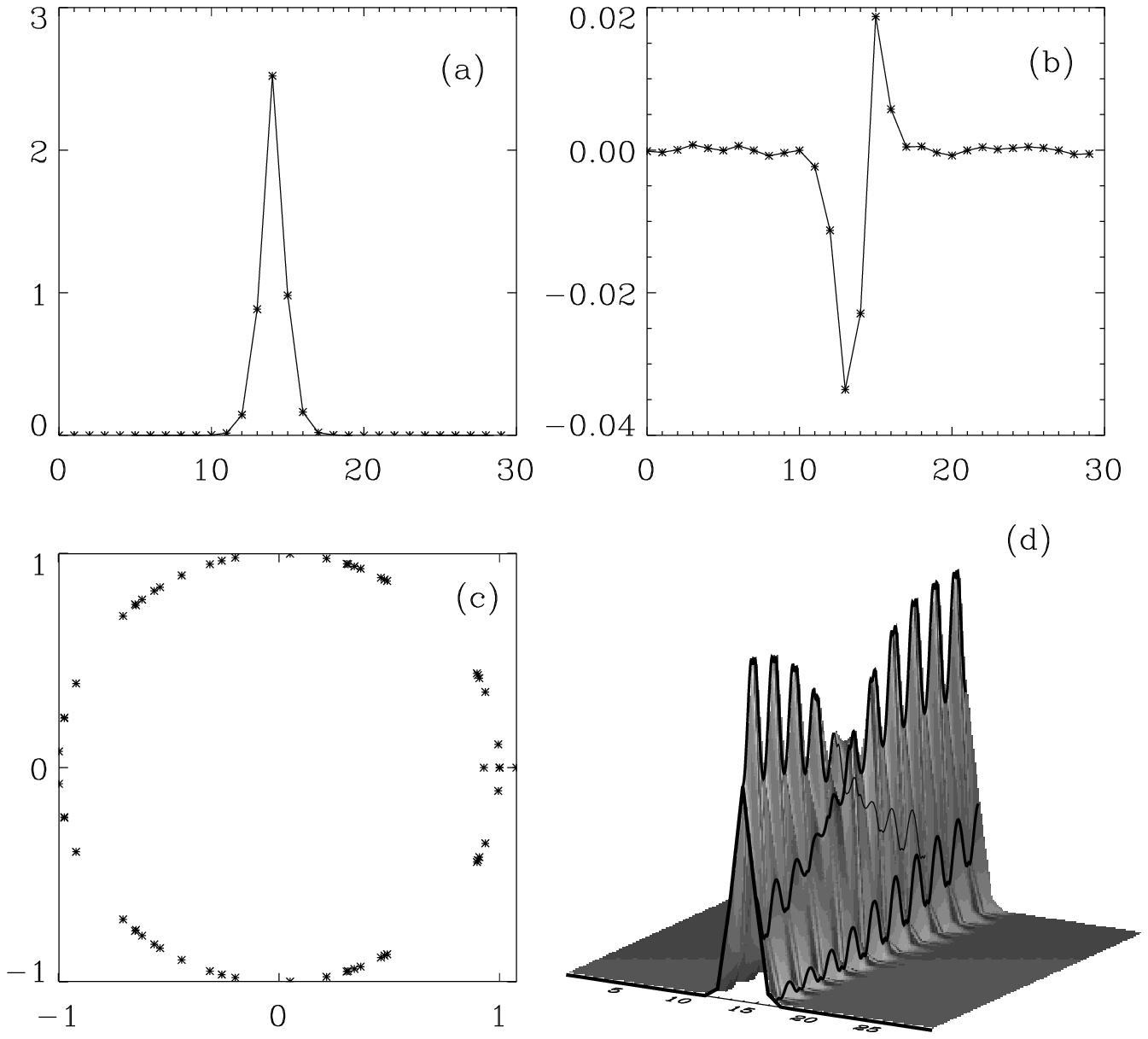} 
\caption{Same as fig.\ref{fig4a} but for
$C=0.1$, $T=92$ and 10 internal oscillations over one period.
Two eigenvalues are out of the unit circle
and this exact solution is linearly unstable.}  \label{fig4b}
\end{figure}

We tested preliminarily the simplest case where $T$ is an integer
multiple of $t_{b}$. For that purpose, we just used the numerical
program initially designed for non time reversible multibreather
solutions \cite{CA97} (a modified Newton method) with few line
adaptations consisting in changing the time periodicity into the skew
periodicity condition

\begin{equation}
\{p_{i}(t),u_{i}(t)\}=\{p_{i+1}(t+T),u_{i+1}(t+T)\}. \label{skewper}
\end{equation}

When a {\em good} approximate solution for a mobile breather can be
obtained (by perturbation of an immobile breather with a translation
mode), it can be used as trying solution for starting the Newton
process. In that case, our program converges quite well to a {\em
numerically exact} mobile breather solution at an accuracy, which is
apparently only limited by the computer precision.  This accuracy has
been pushed till $10^{-21}$ (in quadruple precision) on the skew
periodicity condition 
(\ref{skewper}).

Figs.\ref{fig4a} and \ref{fig4b} show two example of such
solutions. The most striking feature is that the solution does not go
exactly to zero far away from the center of the mobile breather but extends
over the whole system whatever is its size. For the infinite
system, this solution would likely spatially extend to infinity with a
small amplitude oscillating tail.  These mobile breather solutions
look analogous to the nanopteron solutions \cite{FW97} which are
immobile breathers with an infinite phonon tail \footnote{They were
named phonobreather in \cite{Aub97,MA96}. In some limit, their
existence can be proved.}.  The extended Floquet matrix ${\bf F}_{e}$
defined as the derivative of the map $ \{p_{i}(0),u_{i}(0)\}
\rightarrow \{p_{i+1}(T),u_{i+1}(T)\}$ is numerically calculated and
diagonalized in order to look at the stability of its fixed point.

In the case presented in fig.\ref{fig4a}, the mobile breather solution is
linearly stable while in the second case where it is more discrete and
moves slower, the mobile breather is slightly unstable. Further
studies are necessary to explore these phenomena.

\section{Breather Reactivity}

As we found, in subsection \ref{subsection2}, there are marginal modes
at any breather bifurcation but only some marginal modes corresponding
to bifurcations at $\theta=0$ can be used for making a breather
mobile.  What is the effect on the breather of perturbations by
marginal modes which are not associated with pinning modes? Since they
diverge linearly in time, they start a slow transformation of the
breather which is interesting to study at longer time. This
transformation may exhibit complex transitory regimes not yet analyzed
in general. In some cases, it exhibits a simpler behavior which can be
easily interpreted.

 \begin{figure}[tbp] \centering
\resizebox{11cm}{!}{\includegraphics[40pt,420pt][460pt,720pt]{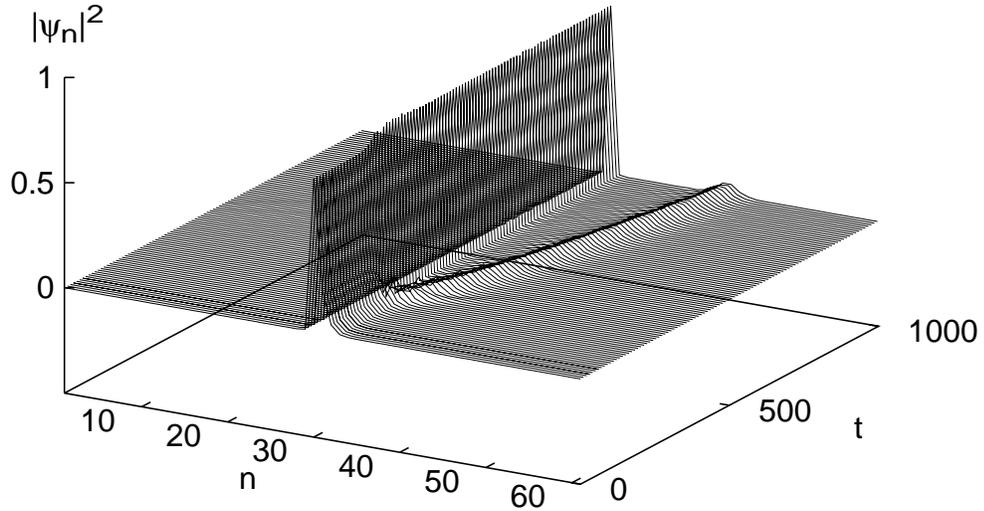}}
\caption{Fission of a 2-breather
state $\ldots\bullet \bullet \bullet 10 \bullet \bullet
\bullet\ldots$ in the DNLS equation by a perturbation by the
marginal mode at the instability threshold ( $C=0.065$,
$\omega_b= (\sqrt{5}-1)/2)$ (from ref.\cite{JA97}).}
\label{fig5}
\end{figure}

For example, for a multibreather (2-breathers bonded state), we
observed that its excitation by a marginal mode may produce a
fission. The most typical case we found (see fig. \ref{fig5}), was
obtained in \cite{JA97} for the DNLS equation
\begin{equation}
	- C (\psi_{n+1}+ \psi_{n-1} -2 \psi_n) - |\psi_n|^2 \psi_n +
	\psi_n = i \dot{\psi}_n \label{DNLS} 
\end{equation}
for the
2-breather state denoted $\ldots\bullet \bullet \bullet 10
\bullet \bullet \bullet\ldots$ which is continued from the
solution at the anticontinuous limit ($C=0$) given by
$\psi_0=\sqrt{1+\omega_b} \e^{i\omega_b t}$ (denoted $1$),
$\psi_1=1$ (denoted $0$) and $\psi_i=0$ for $i\neq 0$ and $1$
(denoted $\bullet$).  There is an instability threshold at
$\theta=0$ at which the marginal mode is time reversible. When
at a given time, a small perturbation colinear to this
marginal mode is added to the breather, this bonded state
breaks after some time into ``Big Brother'' which stay immobile and
``Little Brother'' which is ejected and move quite far away
before stopping. When $C$ is smaller than its value at the
2-breather instability threshold, a critical perturbation
(that is an energy threshold) is needed for having the fission
(Peierls Nabarro energy barrier).  Since the marginal mode is
time reversible, the same behavior occurs when reversing time
so that we get also an elastic collision scheme where running
Little Brother collides elastically with Big Brother.

\begin{figure}[tbp] \centering 
\resizebox{\textwidth}{!}{\includegraphics{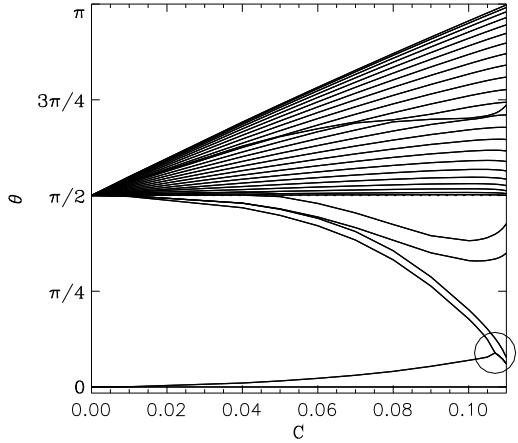}}
\caption{ Arguments $\theta$ of
the eigenvalues of the Floquet matrix ${\bf F}$ versus the
coupling $C$ for the 2-breather $\ldots00-1000100\ldots$ at
$\omega_b=0.8$ of the KG chain with the Lennard-Jones potential.}
\label{fig6a} 
\end{figure}

Let us show another example of perturbation by a marginal mode.
Fig.\ref{fig6a} shows the arguments of the Floquet matrix as a
function of the coupling $C$ for another 2-breather state which also
consist of two identical breathers in anti-phase separated by 3
sites. The lattice is a  KG chain
with the Lennard-Jones on-site potential $V(x)=\frac{1}{72}
(1+(1+x)^{-12}-2(1+x)^{-6})$. This configuration is linearly stable
for small coupling but a Krein bifurcation occurs at $C \approx 0.105$
(the circle on Fig.\ref{fig6a}).
 
\begin{figure}[tbp] \centering 
\includegraphics{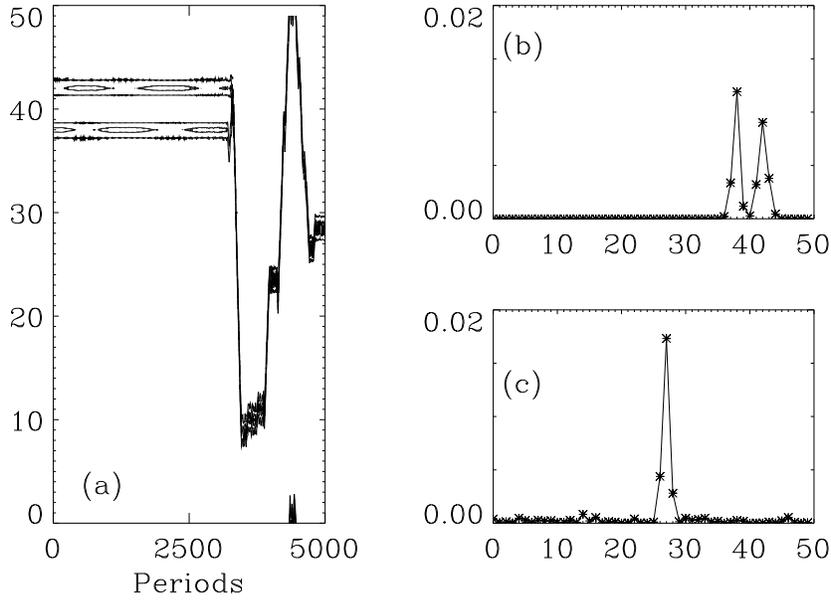}
\caption{Fusion associated with the marginal mode of the Krein
bifurcation shown on fig.\ref{fig6a}.  The norm of the
perturbation is $\lambda \approx 10^{-2}$. (a) is a contour plot of
the energy density versus time, (b) and (c) show the initial and final
energy density.}  \label{fig6}
\end{figure}

A small excitation of this 2-breather by the associated marginal mode
leads, after a long time (about 3000 Periods of the unperturbed breathers,
see fig.\ref{fig6}a) to
a sudden fusion of the excitations. The resulting breather contains
75\% of the total energy and is mobile. Its erratic motion is due to
the radiation emited during the fusion. 

These numerical studies are currently developed.

\section{Concluding Remarks}

In summary, we have shown the existence of a marginal mode growing
linearly in time at any instability threshold. The existence of such a
mode is a necessary condition (although not sufficient) for having a
highly mobile breather and this condition is {\em not} necessarily
related to large spatial breather extension as often believed. This
criteria provides a systematic numerical method for testing breather
mobility and then for finding easily in many models, spatially narrow
breathers which are well mobile.

In addition, an improved Newton method allows to find numerically
``exact'' mobile breathers with an accuracy only limited by the computer
precision.  Actually, these mobile solutions have an infinite tail as
the nanopterons which suggests that generally a strictly localized
breather cannot propagate without radiating energy (but in some case
this radiation can be extremely weak).

The marginal modes, appearing at any instability threshold, do not
necessarily correspond to breather mobility. We found examples where
excitation by a marginal mode produces the fission of a two breather
bonded state.  There are also examples where two colliding breathers
merge into a bigger one (\ie a breather fusion).  Let us
also quote that the inelastic interaction of weak amplitude phonons
with breathers discovered in \cite{CAF97}, can be viewed as a breather
spallation.

Our studies presented here or in the related references, concerns
classical models. However, this series of phenomena are reminiscent of
chemical or nuclear reactions. This should not be surprising when
thinking about the quantized version of these phenomena. Breathers are
bonded (or antibonded) clusters (``molecules'' or ``nuclei'') of $n$
interacting particles which are bosons (physically, these bosons are
``phonons''). Their classical stability depends on the breather
amplitude or frequency, that is in the quantum representation of the
number of bosons bonded together. Thus, depending on the model we may
have stable quantum breathers for some values of $n$ and unstable ones
for other values. Actually, in translationally invariant models, these
breathers form bands but their effective mass may be very high when
they are not mobile. If this breather gets a high mobility, its
effective mass 
should drop drastically\footnote{ We already numerically observed
this phenomena for the quantized bipolarons of the Holstein-Hubbard
model \cite{PA97}.}.  It is also not surprising that these molecule
could react with a chemistry which of course highly depends on the
choice of the boson interactions that is on the anharmonic potentials
of the model.

Finally, let us note that arrays of very resistive Josephson junction 
(which can be modeled by the DNLS equation) are likely the simplest systems
where these ideas on breather propagation and reactivity could be developed
and tested experimentally almost directly.

\begin{ack}
One of us (TC) acknowledges ``la R\'egion Rh\^one-Alpes'' for the grant
``Emergence''. 
\end{ack}


\begin{thebibliography}{99}

\bibitem{SCL73} A.C. Scott, F.Y.F. Chu and D.W. Mc Laughlin,
      {\it proc. IEEE} {\bf 61} (1973) 1443.
      
\bibitem{AL76} M.J. Ablowitz and J.F. Ladik, {\it J. Math. Phys.} {\bf
17} (1976) 1011.

\bibitem{Lan33} L.D. Landau {\it Phys.Z. Sowjetunion} {\bf 3}(1933)  664.

\bibitem{AFKO96} S. Aubry, S. Flach, K. Kladko and E. Olbrich, {\it Phys. Rev. Lett.}
{\bf 76} (1996) 1607-1610.

\bibitem{FMMFA96} L.M. Floria, J.L. Mar\'{\i}n, P.J. Martinez, F. Falo 
and S. Aubry, {\it Europhys. Letts.} {\bf 36} (1996) 539.

\bibitem{Flo97} L.M. Floria et al., this issue.

\bibitem{ST88} A.J. Sievers and S. Takeno {\it Phys.Rev.Letts.} {\bf
61} (1988) 970. 

\bibitem{FW97} S. Flach and C.R. Willis, to appear in {\it Phys. Rep}(1998).

\bibitem{MA94} R.S. MacKay and S.Aubry, {\it Nonlinearity} {\bf 7}
(1994) 1623-1643.

\bibitem{AA90} S. Aubry and G. Abramovici,  {\it Physica} {\bf 43D}
(1990) 199-219. 

\bibitem{Aub94} S. Aubry, {\it Physica} {\bf 71D} (1994) 196-221. 

\bibitem{Aub97}  S. Aubry, {\it Physica} {\bf 103D} (1997) 201-250.

\bibitem{MKS95} R.S. MacKay and J-A. Sepulchre, {\it Physica} {\bf
82D} (1995) 243-254. 

\bibitem{SMK97} J-A. Sepulchre and R.S. MacKay {\it Nonlinearity} {\bf
10} (1997) 1-35. 

\bibitem{FCP97}  K. Forinash, T. Cretegny and M. Peyrard, {\it
Phys. Rev.} {\bf E55} (1997) 4740. 

\bibitem{AP97} S. Aubry and M. Peyrard, in preparation.

\bibitem{LSMK97} R. Livi, M. Spicci, R.S. MacKay, {\it Nonlinearity}
{\bf 10} (1997) 1421-1434.

\bibitem{MK97} R.S. MacKay, this issue (1998).

\bibitem{CA97} T. Cretegny and S. Aubry, {\it Phys. Rev.}{\bf B55}
(1997) R 11929-32.

\bibitem{MA96} J.L. Mar\'{\i}n and S.Aubry, {\it Nonlinearity} {\bf 9}
(1996) 1501-1528.

\bibitem{CAT96}  Ding Chen, S. Aubry and G. Tsironis, {\it
Phys.Rev.Letts} {\bf 77} (1996) 4776-4779. 

\bibitem{JA97} M. Johansson and S. Aubry, Nonlinearity {10} (1997) 1151-1178.

\bibitem{Fle97} V. Fleurov et al., this issue.

\bibitem{HT92} K. Hori and S. Takeno, {\it J.Phys.Soc.Japan} {\bf 
61} (1992) 2186  and 4263.

\bibitem{SPS92} K.W. Sandusky, J.B. Page and K.E. Schmidt {\it Phys.Rev.}
{\bf B46} (1992) 6161.

\bibitem{BP95}  O. Bang and M. Peyrard, {\it Physica} {\bf 81D} 
(1996) 433.

\bibitem{FW95} S. Flach and C.R. Willis, {\it Phys.Rev.Lett.} {\bf 72} (1994) 1777.

\bibitem{Aub96} S. Aubry, Oral Comm. in Heraklion,
Greece,Sept.30-Oct.4 (1996), unpublished.

\bibitem{CMA97} T. Cretegny, J.L. Mar\'{\i}n and S.Aubry, in preparation.

\bibitem{CAF97} T. Cretegny, S. Aubry and S.Flach, this issue.

\bibitem{AA68} V.I. Arnold, A.Avez {\it Ergodic Problems of Classical 
Mechanics} App.29, W.A. Benjamin Inc. (1968)

\bibitem{PA97} L. Proville and S. Aubry, Proceeding of 
{\it Fluctuations, Nonlinearity and Disorder} (Heraklion,
Greece,Sept.30-Oct.4 1996) ed.  G.P.  Tsironis, to be psublished in
Physica D (1998) and in preparation.

\bibitem{Fla97} S. Flach et al, this issue.

\end{thebibliography}
\end{document}